\documentclass[a4paper,twoside]{article}

\usepackage{epsfig}
\usepackage{subcaption}
\usepackage{calc}
\usepackage{amssymb}
\usepackage{amstext}
\usepackage{amsmath}
\usepackage{amsthm}
\usepackage{multicol}
\usepackage{pslatex}
\usepackage{apalike}

\usepackage{graphicx}
\usepackage[hang,flushmargin]{footmisc} 
\graphicspath{ {./images/} }
\usepackage{subcaption}
\usepackage{amsthm}
\newtheorem{protocol}{Protocol}[section]

\usepackage{SCITEPRESS}     

\begin{document}
\renewcommand{\thefootnote}{\fnsymbol{footnote}}
\title{LASER: Lightweight And SEcure Remote keyless entry protocol (Extended version)$^*$\footnotemark[0]}

\author{\authorname{Vanesa Daza and Xavier Salleras}
\affiliation{Department of Information and Communication Technologies \\
Universitat Pompeu Fabra, Barcelona, Spain \\
CYBERCAT - Center for Cybersecurity Research of Catalonia}
\email{\{vanesa.daza, xavier.salleras\}@upf.edu}
}


\abstract{Since Remote Keyless Entry (RKE) systems started to be widely used, several vulnerabilities in their protocols have been found. Attacks such as jamming-and-replay attacks and relay attacks are still effective against most recent RKE systems \cite{Ibrahim2018HackingKeyless}, even when many secure schemes have been designed. Although they are interesting from a theoretical point of view, the complexity of these solutions is excessive to implement them into a fob \cite{bluetoothIssues}. This paper presents a lightweight and general solution based on a one message protocol, which guarantees the integrity and validity of the authentication in RKE systems, protecting the communication against the well-known jamming-and-replay and relay attacks, without using complex cryptographic schemes. Moreover, we also adapt our protocol for passive RKE (PRKE) systems. Our solution also includes a novel frequency-hopping-based approach which mitigates deny-of-service attacks. Finally, a prototype has been implemented using non-expensive hardware. Obtained results assure scalability, effectiveness and robustness.} 

\onecolumn
\maketitle
\normalsize 
\vfill

\section{\uppercase{Introduction}}
\label{sec:introduction}
\renewcommand{\thefootnote}{\fnsymbol{footnote}}

\footnotetext[1]{This is an extended version of a paper by the authors published in Proceedings of SECRYPT 2019.}

\renewcommand{\thefootnote}{\arabic{footnote}}
\setcounter{footnote}{0}

\noindent
The usage of RKE systems has been increasing over the years, being them widely used to remotely lock and unlock cars, garage doors, sensors, doorbells or alarms. The first RKE systems used a simple protocol, where a code was sent in plaintext to a receiver which had to execute a command, let us say, unlock a door. However, as sniffing and replaying the code was enough to be able to unlock such a door, a new scheme called \textit{rolling codes} was developed, and it is still widely used nowadays. Such scheme pretends to be secure so the key fob computes and sends a new code each time it is used, and each code is accepted by the receiver just once. Even so, it has been proved that rolling codes are vulnerable to different attacks, and authorities are starting to report\footnote{https://www.west-midlands.police.uk/news/watch-police-release-footage-relay-crime} criminals taking profit of these vulnerabilities. This fact has led researchers to design new secure schemes \cite{6201976} to protect these systems, but their complexity made manufacturers not to implement them, so it would mean to develop key fobs with some disadvantages, i.e. a higher price or a faster draining of the battery. This is due to the fact that many solutions proposed to use cryptographic schemes \cite{aes2007} which needed higher computing power than the available in the current fobs. Furthermore, the proposed protocols usually need more than one message to exchange some private information or instruction command. For example, some solutions \cite{7921990} require to use a 4-way handshake before sending an instruction command, which increases the complexity of the protocol.

\textbf{Contributions.} We provide a secure protocol\footnote{The presented solution has been submitted as an invention to be patented with European Patent application number 19382339.0, on May 6th, 2019.} to be implemented by manufacturers into both RKE and PRKE systems. Our scheme is robust against both jamming-and-replay attacks and relay attacks; furthermore, it mitigates the effectiveness of jamming-based deny-of-service attacks, thanks to the integration into the protocol of a frequency-hopping approach. Moreover, our solution is a one message protocol for RKE systems and a two messages protocol for PRKE systems, where both approaches use a hash function proved to have low CPU resources consumption. As such, our solution is lightweight, scalable and easy-to-implement. The purpose of this solution is to be applied into key fobs with the only requirement of having a real-time clock, synchronized periodically as detailed in our protocol. We also demonstrate how our solution can be implemented achieving good results.

\textbf{Paper organization.} This paper is structured as follows: In Section \ref{sec:background} we explain both RKE and PRKE systems along with the common attacks that can be performed against them. In Section \ref{sec:relatedwork} the state-of-the-art is presented. In Section \ref{sec:ourscheme} we explain our solution. The implementation of the proposed solution is described in Section \ref{sec:implementation}. The experiments and the results derived from them are explained in Section \ref{sec:experiments}. We finally conclude in Section \ref{sec:conclusion}.

\section{\uppercase{Background}}
\label{sec:background}
\noindent
In this section we first explain both RKE and PRKE systems, as well as the main protocols that are currently used. Later, we explain the main attacks that are currently effective against them.

\subsection{Remote Keyless Entry systems}

We call RKE to those systems which are composed of a fob \textit{F} and a device \textit{D}. When a button on \textit{F} is pressed, a radio frequency signal is sent to \textit{D}, including an instruction command that \textit{D} will have to execute. These systems are commonly used to lock or unlock cars and open their boots, to open a garage door, to control a temperature sensor, etc. The main protocols used by these systems can be divided as follows:

\begin{itemize}
\item \textit{Fixed codes.} This is the simplest scheme. As depicted in Figure \ref{subfig:simpleprotocol}, \textit{F} sends a command \textit{cmd} to \textit{D}, which is essentially a bit stream referring to an action that \textit{D} will have to perform.

\begin{figure*}%
\centering
\begin{subfigure}{.67\columnwidth}
\includegraphics[width=\columnwidth]{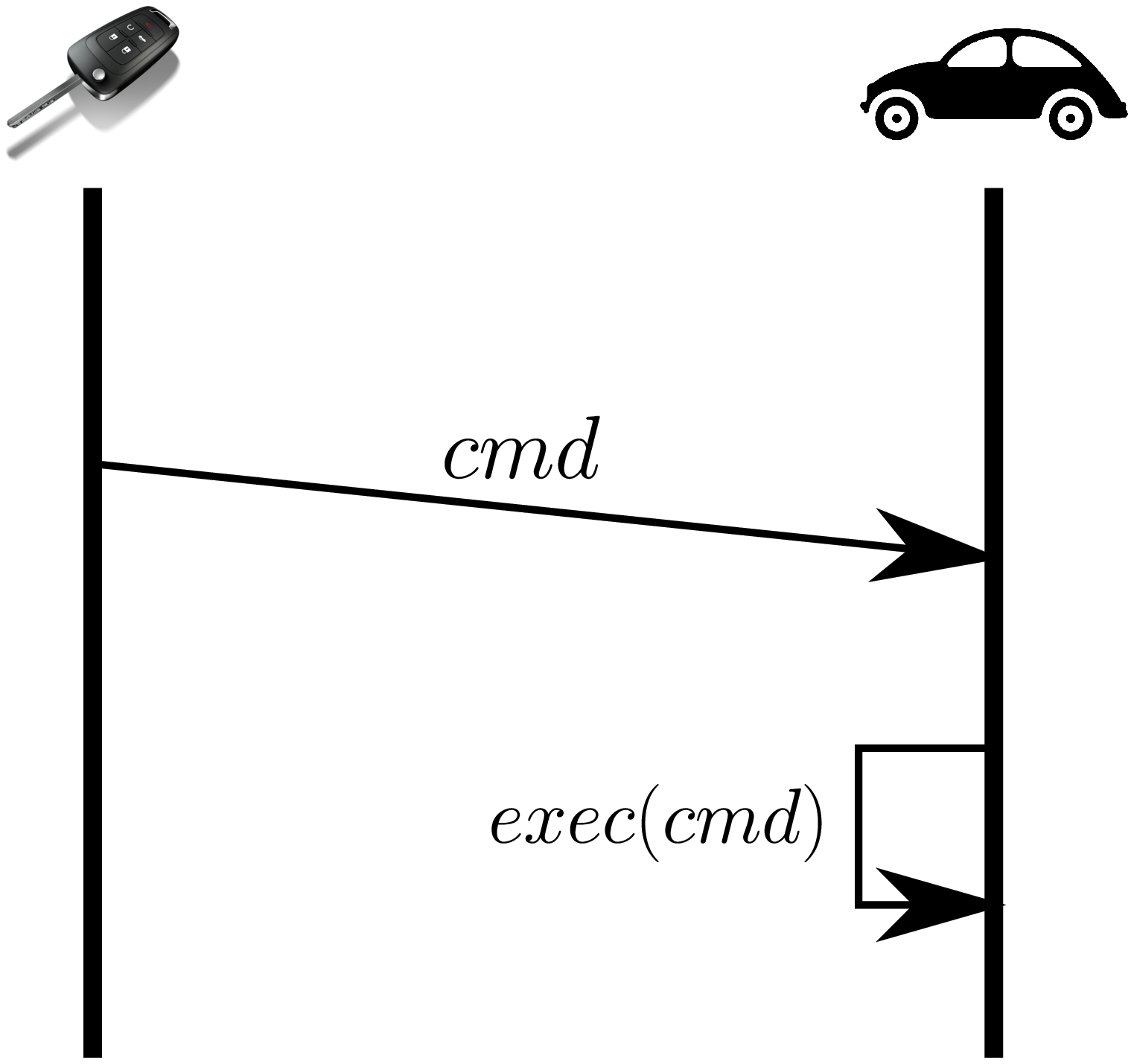}%
\caption{Fixed codes protocol}%
\label{subfig:simpleprotocol}%
\end{subfigure}\hfill%
\begin{subfigure}{.67\columnwidth}
\includegraphics[width=\columnwidth]{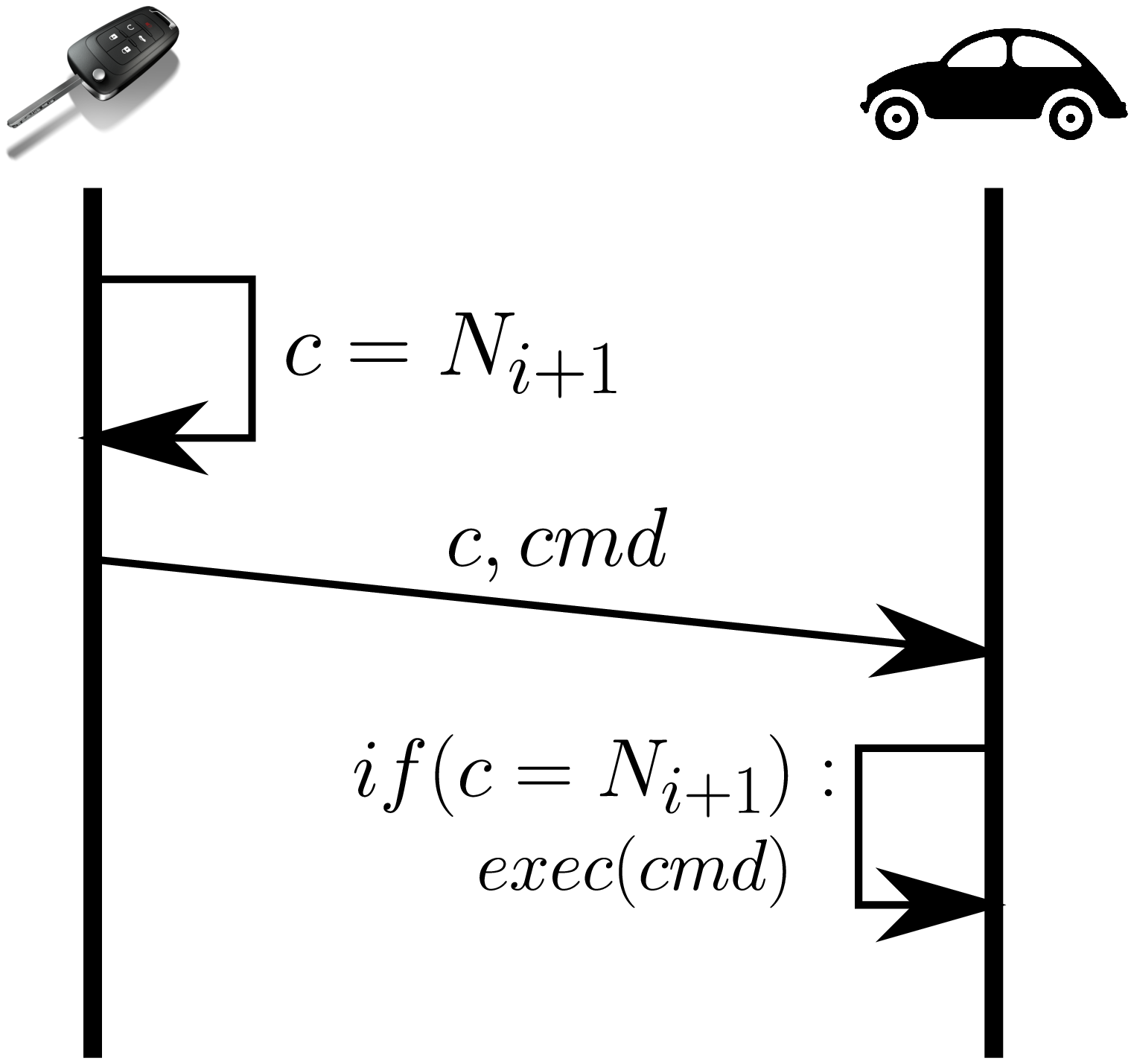}%
\caption{Rolling codes protocol}%
\label{subfig:rollingprotocol}%
\end{subfigure}\hfill%
\begin{subfigure}{.67\columnwidth}
\includegraphics[width=\columnwidth]{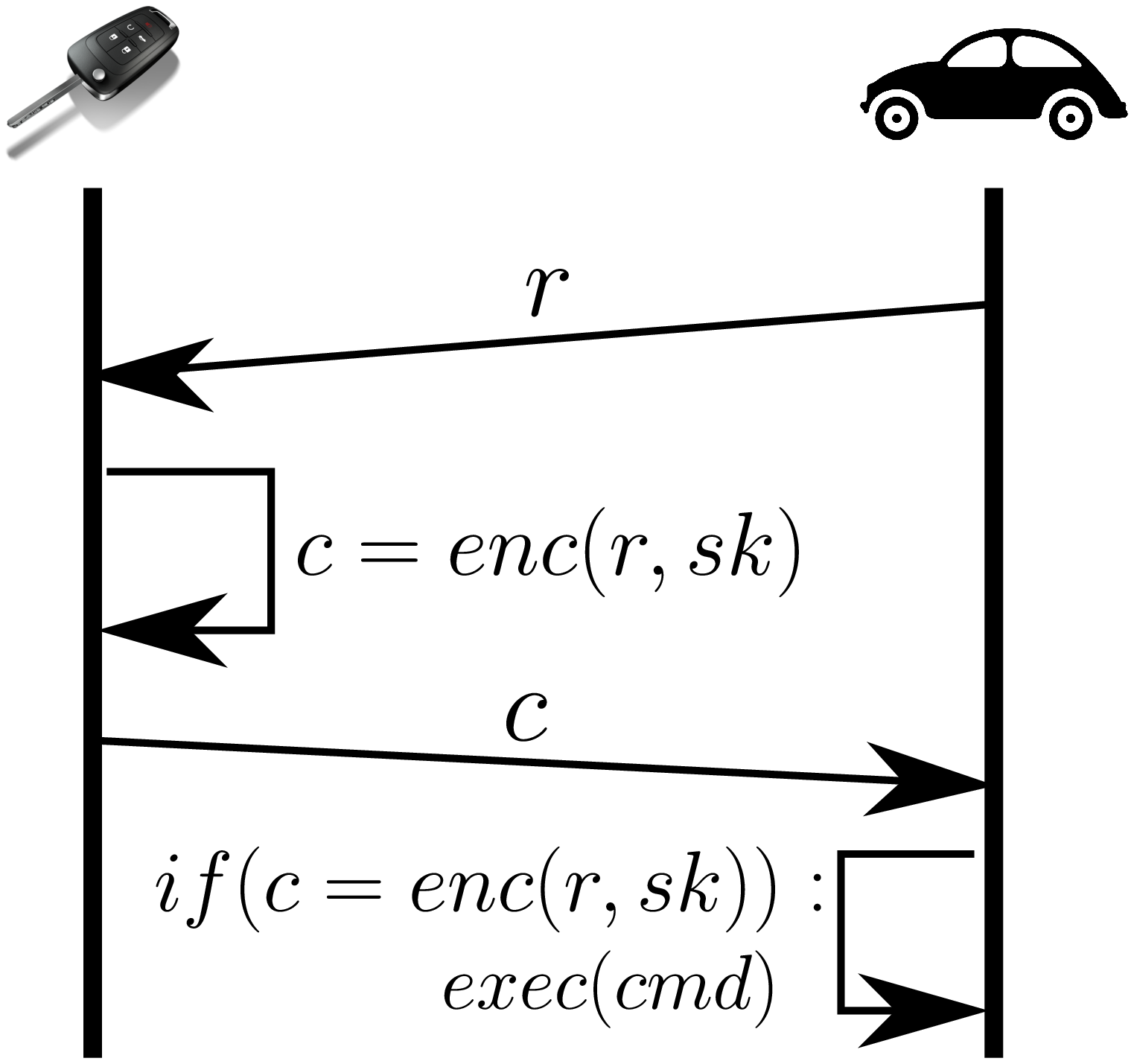}%
\caption{Challenge-response protocol}%
\label{subfig:challengeprotocol}%
\end{subfigure}%
\caption{Main RKE and PRKE protocols}
\label{fig:mainkesprotocols}
\end{figure*}

\item \textit{Rolling codes.} There is a wide variety of rolling codes algorithms, but all of them rely on the idea of sending different codes each time a button of \textit{F} is pressed. In order to accomplish this purpose, both \textit{F} and \textit{D} have previously agreed on a secret key from which derives a sequence of codes $N_1,N_2,...,N_p$. Then, as depicted in Figure \ref{subfig:rollingprotocol}, each time a button on \textit{F} is pressed, the next code \textit{c} is computed and sent to \textit{D}, who checks if the received number is equal to a value \textit{c} that previously it also computed. Apart from \textit{c}, a command \textit{cmd} is also sent, which is typically a sequence of bits that refers to an action \textit{D} will have to do, i.e. unlock a car. Each value \textit{c} can be used only once. In case \textit{D} may have not received some of the codes sent by \textit{F}, it commonly checks up to the next 256 generated codes, and when a correct value \textit{c} is received by \textit{D}, all the codes behind it cannot be used again. One of the most used rolling codes devices has been KeeLoq \cite{KeeLoq}. 

\end{itemize}

\subsection{Passive Remote Keyless Entry systems} 

Passive Remote Keyless Entry (PRKE) systems \cite{prke} are a special type of RKE. PRKE systems do not require the user to manipulate \textit{F}. Instead, as soon as \textit{D} receives an external input (i.e. if \textit{D} is a door, someone pulling the handle), it automatically sends a request to \textit{F}, which replies with a confirmation. The most used protocol \cite{challengeres} for PRKE systems is the challenge-response protocol: 

\begin{protocol}[Challenge-response protocol for PRKE]
\label{prot:challres}
Both \textit{D} and \textit{F} perform the following 2-message handshake:
\begin{enumerate}
  \item First, \textit{D} computes a random value \textit{r} (the challenge), and sends it to \textit{F}.
  \item \textit{F} encrypts \textit{r} using a pre-shared symmetric-key $sk$, and sends the encrypted value \textit{c} to \textit{D}.
  \item \textit{D} decrypts \textit{c} using the same key $sk$ and verifies the identity of \textit{F} .
\end{enumerate}
\end{protocol}

For example, if \textit{D} is a car using a challenge-response protocol, when the user carrying \textit{F} pulls the car handle, \textit{D} sends a message with a challenge \textit{r}, and as soon as \textit{F} receives it, it replies with its answer. This is depicted in Figure \ref{subfig:challengeprotocol}.

\subsection{Attacks against RKE and PRKE}

\begin{figure}[ht]
    \centering
    \includegraphics[width=0.42\textwidth]{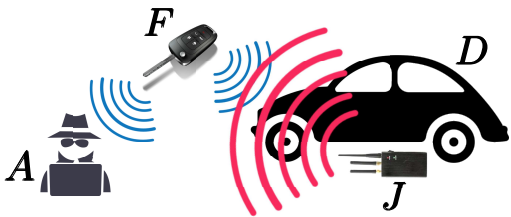}
    \caption{Jamming-and-replay attack}
    \label{fig:jamm}
\end{figure}

\textbf{Jamming-and-replay attack.} As depicted in Figure \ref{fig:jamm}, these attacks \cite{jammingreplay} are performed using two transceiver devices. One of them is placed near to \textit{D}, hidden from the view of the victim \textit{V}, and jamming the frequency used by the system an attacker \textit{A} is willing to hack. Then, the other one is close to \textit{F}, eavesdropping the communications. When \textit{V} presses the button of \textit{F}, the signal it sends is jammed by the jamming transceiver \textit{J}, and \textit{V} is forced to use an alternative (i.e. a physical key). Meanwhile, \textit{A} captures the message sent by \textit{F}, and as \textit{D} never receives it, \textit{A} will be able to replay it later. Finally, the jammer can be remotely deactivated by \textit{A}, as soon as he is sure that \textit{V} will not try to use \textit{F} again.

\begin{figure}[ht]
    \centering
    \includegraphics[width=0.45\textwidth]{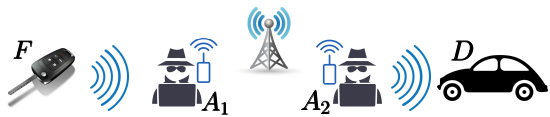}
    \caption{Relay attack}
    \label{fig:relay}
\end{figure}

\textbf{Relay attack.} As it can be seen in Figure \ref{fig:relay}, this kind of attacks \cite{cryptoeprint:2010:332} are performed using two transceivers connected through an LTE network or similar. One of them is close to \textit{D}, and the other one to \textit{F}. Like this, they create a bridge between both endpoints. If the attacked system is a PRKE, when the attacker $A_2$ pulls the car handle the challenge-response protocol is performed through the bridge created by both attackers. Otherwise, if we are talking about an RKE system, we have to expect that the user may either accidentally press the button on \textit{F}, or leave it unattended (thus allowing the attacker $A_1$ to press the button).

\textbf{Deny-of-service (DoS) attack.} This kind of attack \cite{jamming} is also based on jamming the frequency used by the protocol, but in this case with the main goal of denying the service. It has a lower impact on the system security as it does not grant access to the system, but it bothers the user, who will require a physical key if he wants to perform the action.

\section{\uppercase{Related work}}
\label{sec:relatedwork}
\noindent
Regarding the attacks against RKE systems, an important contribution on the topic has been recently done in \cite{Ibrahim2018HackingKeyless}. They demonstrate as the jamming-and-relay attacks are nowadays still effective against a wide variety of modern cars, by making use of two units of a radio frequency device called HackRF One\footnote{https://greatscottgadgets.com/hackrf/}, one for jamming and the other one for logging data and replaying later.

A particular RKE scheme based on rolling codes, and widely used by many manufacturers, is called \textit{Hitag2} \cite{hitag2}. An important contribution related to this type of RKE has been done in \cite{lockit}, where a novel correlation-based attack is presented. This attack allows an attacker to recover the secret key used in \textit{Hitag2} systems, just by eavesdropping at least four of the codes sent by the fob. Thus, it allows the attacker to clone the fob. As stated in the paper, major manufacturers have sold systems with this vulnerability for over 20 years. As such, the need for new secure and easy-to-implement schemes becomes clear.

\textbf{Implementation of the attacks.} By making use of two radio frequency devices called Yardstick One\footnote{https://greatscottgadgets.com/yardstickone/} (YS1), a jamming-and-replay attack can be performed by using a python implementation\footnote{https://github.com/exploitagency/rfcat-rolljam} of this attack. This implementation makes use of a library called \textit{rflib}, included in a software used by YS1 called RfCat\footnote{https://github.com/atlas0fd00m/rfcat}. That said, one antenna will be jamming while the other will be sniffing the code of the fob. The same implementation is useful for performing just the DoS attack. Moreover, taking this implementation as a starting point, implementing a relay attack is trivial.

\textbf{Proposed solutions.} Many secure schemes \cite{6201976}, \cite{7921990} have been designed to increase the security of RKE and PRKE systems. The main problem they present is their complexity, so they use cryptographic schemes which are hard to implement into cheap key fobs. On the other hand, some schemes \cite{8320107} have been proved to be both simple and effective against relay attacks. One of them, proposed in \cite{7950848}, demonstrates that a protocol calculating the time between message exchanges can determine if a relay attack is being performed against a PRKE or not. This is the main idea behind \textit{LASER}, which also solves the replay vulnerability.

\section{\uppercase{Our scheme: LASER}}
\label{sec:ourscheme}
\noindent
In this section we explain step-by-step our protocol, \textit{LASER}, for both RKE and PRKE systems. We consider a fob \textit{F} and a generic device \textit{D}, assuming it to be a car. First, both endpoints have to agree on a randomly generated secret key \textit{sk} large enough to make a brute-force attack hard to accomplish (i.e. a 256-bits key). They also need to agree on a set of commands \textit{cmd}, used for example to lock the car, unlock it, etc. \textit{D} also has a car identification number ($device_{id}$) known by \textit{F}. 

In both RKE and PRKE systems, both \textit{F} and \textit{D} will be required to compute a hash. The hash function used by both devices was required to be lightweight in order to optimize the timings and the resources consumption. For our implementation and analysis we have chosen to use \textit{Blake2}, a hash function proposed in \cite{blake2}, which guarantees a low power and computing resources consumption. Furthermore, it is proved to be as fast as \textit{MD5}, but solving the security vulnerabilities \textit{MD5} presents. 

In particular we are interested in using \textit{Blake2s}, a version of \textit{Blake2} optimized for 8-bit platforms, which are the kind of cheap processors commonly used for key fobs. Basing our solution in the usage of a hash function like \textit{Blake2} instead of using some complex cryptographic scheme, we are decreasing the costs of implementing our solution, and also avoiding a fast draining of the battery.

Our solution performs a frequency-hopping protocol where the frequency channel used to transmit the messages changes each period of time $p$. This means that both \textit{D} and \textit{F} must agree on the same channel, and to achieve it they perform the Protocol \ref{prot:freqhop}.

\theoremstyle{definition}
\begin{protocol}[Frequency-hopping for LASER]
\label{prot:freqhop}
The frequency-hopping for a specific endpoint, which has a number of available frequency channels $N_c$, is performed as follows:
\begin{enumerate}
  \item Each period of time \textit{p} (both F and D have previously agreed on this value) it gets the current datetime in a timestamp form, sums the secret key \textit{sk} to it and calculates its hash \textit{h}.
  \item It calculates the channel $ch$, which is the modulo $N_c$ of the integer representation of \textit{h}: $ch \equiv int(h) \pmod {N_c}$.
\end{enumerate}
\end{protocol}

Next subsections explain the specific details for either RKE and PRKE systems.

\subsection{LASER for RKE}

In this subsection we first explain all the steps of the RKE protocol in detail, and later the main approach used to prevent each kind of attack.

\subsubsection{LASER for RKE: protocol details}

\begin{figure}[ht]
    \centering
    \includegraphics[width=0.40\textwidth]{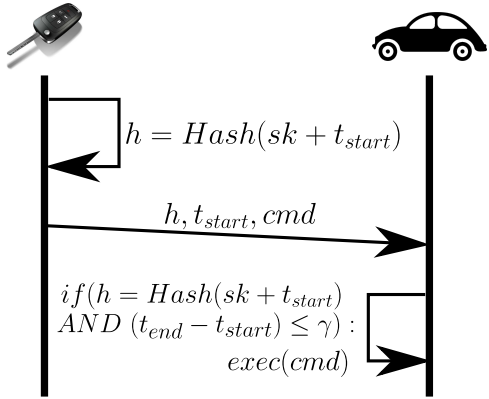}
    \caption{LASER for RKE}
    \label{fig:protocol}
\end{figure}

In this scheme, \textit{D} is required to be always listening to a specific channel, so it will be continuously performing the Protocol \ref{prot:freqhop}. However, \textit{F} will perform it just before to start the Protocol \ref{prot:activescheme}. When the owner of \textit{D} wants to execute a command \textit{cmd} by pressing a button on \textit{F}, \textit{F} calculates $ch$ by first calculating $h$, but rounding the timestamp to the previous multiple of $p$. Then, next protocol is performed (as depicted in Figure \ref{fig:protocol}):

\begin{protocol}[LASER for RKE]
\label{prot:activescheme}
Both \textit{D} and \textit{F} follow the next protocol:
\begin{enumerate}
  \item \textit{F} takes the current timestamp $t_{start}$, sums it to \textit{sk} and computes its hash \textit{h}.
  \item \textit{F} sends \textit{h} over $ch$ along with the real timestamp $t_{start}$ and the command cmd.
  \item As soon as \textit{D} receives the message sent in the last step, it gets the current timestamp $t_{end}$, and checks if the difference between $t_{start}$ and $t_{end}$ is lower than or equal to a threshold $\gamma$, previously estimated.
  \item If the above condition is true, and $h$ is correct, \textit{D} executes \textit{cmd}.
\end{enumerate}
\end{protocol}

An accurate time synchronization between $F$ and $D$ is crucial, as $F$ has to send an exact timestamp. To overcome this drawback, we propose the usage of the same approach we introduced in our protocol: if $F$ sends a timestamp $t_{start}$ that does not verifies $(t_{end} - t_{start}) \leq \gamma$, $D$ replies with a message $h_{sync},t_{sync}$, where $t_{sync}$ is the correct timestamp and $h_{sync} = Hash(sk + t_{sync})$. $F$ updates its real-time clock after verifying $h_{sync}$. The purpose of sending also a hash here, is to avoid an attacker being able to send messages to $F$ to modify its current time.

\subsubsection{LASER for RKE: security analysis}

\textbf{Preventing jamming-and-replay in RKE.} To prevent jamming-and-replay, our solution sends a unique hashed value \textit{h} of a string. Such string results of concatenating a secret key \textit{sk} and the current timestamp $t_{start}$ at the moment the protocol is initiated. Like this, each hash will be unique in time, and will be accepted by the receiver just at that moment. Plus, the fact of concatenating a secret key makes impossible for an attacker \textit{A} to generate a new hash.

\textbf{Preventing relay attack in RKE.} We first need to estimate the threshold $\gamma$, which is the maximum amount of time a message should take going from \textit{F} to \textit{D}. In this scenario, if a message took an amount of time ($t_{end} - t_{start}$) higher than $\gamma$, we could say that \textit{F} is placed further from \textit{D} than what it should be, and that the protocol is performed by means of a relay attack, using an LTE network or similar.

\textbf{Preventing DoS in RKE.} Both endpoints have a range of frequency channels $N_c$ available to perform the frequency-hopping protocol, and the aim is to agree on a channel $ch$ without an attacker being able of knowing it. The purpose is to change the transmitting channel each short period of time \textit{p} (let us say, 10 seconds), which should be defined by the manufacturer considering the best performance of the device. By doing this, an attacker willing to perform a DoS attack against us will have to jam a wide range of frequencies at the same time. It can be done by means of several jamming devices, which is an expensive investment\footnote{https://www.jammer-store.com/hpj16-all-frequencies-jammer.html}.

\subsection{LASER for PRKE}

In this subsection we first explain all the steps of the PRKE protocol with detail, and later we introduce the main approach used to prevent each kind of attack.

\subsubsection{LASER for PRKE: protocol details}

\begin{figure}[ht]
    \centering
    \includegraphics[width=0.38\textwidth]{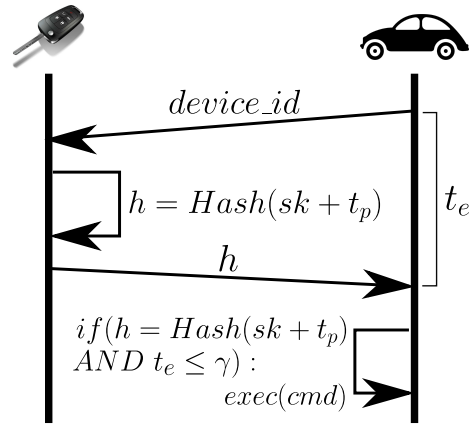}
    \caption{LASER for PRKE}
    \label{fig:protocolp}
\end{figure}

In this scheme, it will be \textit{F} who is continuously performing the Protocol \ref{prot:freqhop}. When the owner of \textit{D} wants to unlock it by pulling the handle, \textit{D} calculates $ch$ by first calculating $h$, but rounding the timestamp to the previous multiple of $p$. Then, next protocol is performed (as depicted in Figure \ref{fig:protocolp}):

\begin{protocol}[LASER for PRKE]
\label{prot:passivescheme}
Both \textit{D} and \textit{F} follow the next protocol:
\begin{enumerate}
  \item \textit{D} sends over $ch$ a \textit{SYN} message to \textit{F} including the $device_{id}$. At the moment it sends the message, it also starts to calculate a message exchanging time $t_e$.
  \item \textit{F} computes the hash value of \textit{sk} plus $t_p$, and sends the result \textit{h} to \textit{D}.
  \item As soon as \textit{D} receives the message sent in the last step, it stops the counter of $t_e$. Like this, now \textit{D} knows a value $t_e$ which is the time between \textit{D} sending a message and receiving a response. If the received value \textit{h} is correct and $t_e$ is lower than or equal to a threshold $\gamma$, \textit{D} executes the desired action.
\end{enumerate}
\end{protocol}

In PRKE, if $D$ does not receive a response after sending the first message of the protocol, it can be that $t_p$ on $F$ is incorrect. In this case, $D$ must send $h_{sync},t_{sync}$ using all the other frequencies, to be able to reach the one used by $F$, and make it update its current time.

\subsubsection{LASER for PRKE: security analysis}

\textbf{Preventing jamming-and-replay in PRKE.} To prevent jamming-and-replay, in PRKE we also send a unique hashed value \textit{h} of a string. Although we also compute \textit{h} concatenating \textit{sk} and a timestamp, in this case the later is slightly different. For PRKE the prevention against relay attacks is based on another approach we explain in next paragraph, and this is the reason why we can use the timestamp $t_p$ calculated during the frequency hopping protocol as the value concatenated to \textit{sk}. Like this, each \textit{h} can be used only during a short period of time \textit{p}, thus preventing jamming-and-replay.

\textbf{Preventing relay attacks in PRKE.} The value $t_e$ is the time it takes a message to go from $D$ to \textit{F}, plus a response message to go back to \textit{D}. By placing \textit{F} next to \textit{D} and pulling the handle of the car, we can calculate an estimated value $\gamma$, which is the threshold the protocol should never surpass. If a message took an amount of time $t_e$ higher than $\gamma$, we could say that \textit{F} is placed further from \textit{D} than what it should be, and that the protocol is performed by means of a relay attack. As in this case is \textit{D} who calculates $t_e$, \textit{F} will not be required to calculate the current timestamp $t_{start}$, thus the protocol will be less time and power consuming for it.

\textbf{Preventing DoS in PRKE.} For PRKE systems, the prevention against DoS attacks works essentially like in RKE systems.

\section{\uppercase{Implementation}}
\label{sec:implementation}
\noindent
Our solution has been implemented using the following hardware:

\begin{itemize}
    \item A PC with a CPU Intel Core i5 3210M and Kali Linux installed, representing the device \textit{D}.
    \item A PC with a CPU Intel Atom x7-z8750 and Kali Linux installed, representing the fob \textit{F}.
    \item Two units of YS1: one plugged in the PC representing \textit{D} and the other one plugged in the PC representing \textit{F}.
\end{itemize}

This prototype\footnote{https://github.com/xevisalle/laser} has been developed and tested with both PCs having Python, RfCat and some other dependencies installed. Next subsections explain the details of this implementation for both RKE and PRKE systems.

\subsection{Implementation of LASER for RKE}

Our code is composed of a single script, which can be run either for \textit{F} and \textit{D}. Once it has been run in the first PC (acting as \textit{D}) providing a $device_{id}$ and a $sk$, it starts performing the frequency-hopping protocol. The range of frequencies used by the code can be provided by the user, where the available range depends on the antenna used, in this case YS1. While performing the frequency-hopping protocol, it also starts to listen for incoming messages. In the case of \textit{F}, the script will remain waiting for a user input, which will be the command to be executed. 

Once we press the key corresponding to the command we want to execute, \textit{F} will first perform the frequency-hopping protocol to determine which frequency has to use, and after this, it will perform the \textit{LASER} protocol. The message sent by \textit{F} will be like the one depicted in Figure \ref{fig:msgrke}, where $start$ and $end$ are 4 always identical bytes placed at the beginning and at the end of the message, to make it easier for the receiver to catch it. Furthermore, the protocol keeps being performed while the user retains the key pressed, meaning this that messages are sent continuously till the key is released. Even pressing and releasing the key quickly, our tests demonstrate that around 6 messages are sent on average. This has been done on purpose, like is done in regular RKE systems, to avoid having to press more than once if the receiver is not able to catch the message the first time, due to random hardware errors. Finally, once \textit{D} receives and verifies the message hash and the timestamps, it executes the command.

\begin{figure}[ht]
\resizebox{\columnwidth}{!}{%
\begin{tabular}{|c|c|c|c|c|c|}
  \hline
  start & device\_id & hash & t\_start & cmd & end \\
  4 bytes & 4 bytes & 6 bytes & $\sim$ 20 bytes & 2 bytes & 4 bytes \\
  \hline
\end{tabular}%
}
\caption{LASER message for RKE}
\label{fig:msgrke}
\end{figure}

\subsection{Implementation of LASER for PRKE}

We run the code like we did with the one for RKE. In this case, \textit{D} expects a user input which simulates, for example, pulling the car handle. Once done, it performs the frequency hopping protocol to know at which frequency \textit{F} is expecting to receive messages. After this, it sends a message like the one depicted in Figure \ref{fig:msgprke}, where $hash$ is null. As it happens with the implementation of \textit{LASER} for RKE, \textit{D} keeps sending messages while is receiving the input from the user (i.e. while the user is pulling the car handle), in order to guarantee the performance of the protocol.

\begin{figure}[ht]
\centering
\begin{tabular}{|c|c|c|c|}
  \hline
  start & device\_id & hash & end \\
  4 bytes & 4 bytes & 6 bytes & 4 bytes \\
  \hline
\end{tabular}
\caption{LASER message for PRKE}
\label{fig:msgprke}
\end{figure}

Once \textit{F} (who was performing the frequency-hopping protocol as well) receives the message from \textit{D}, it replies with a new message, this time with the $hash$ field filled. \textit{D} receives the message and after verifying its hash and the timestamps, it executes the specified command \textit{cmd}.

\section{\uppercase{Experiments and results}}
\label{sec:experiments}
\noindent
In this section we estimate the threshold $\gamma$, and then we use it to analyze the robustness of both systems against relay attacks.

\subsection{Estimating the threshold}

For each system RKE and PRKE we have tried to execute a command one thousand times. The success rate has been 100\% in both cases, meaning this that the command has been always executed. By logging the timestamps into a dataset, we have found out that the time it takes for a message to go from an endpoint to the other one is never higher than $t_{max} = 136 \ ms$ for the RKE solution, as shown in Table \ref{tab:ev}. For PRKE systems, where the calculated time is how much it takes \textit{D} to receive \textit{F}'s reply, the maximum time it took has been $t_{max} = 175 \ ms$. \\

\begin{table}[h]
\vspace{-0.2cm}
\caption{Information extracted from timestamps of RKE and PRKE systems, expressed in milliseconds.}\label{tab:ev} \centering
\begin{tabular}{|c|c|c|c|c|}
  \hline
  System & $t_{max}$ & $t_{min}$ & $t_{avg}$ & $t_{Q3}$ \\
  \hline
  RKE & 136 & 55 & 71 & 79 \\
  \hline
  PRKE & 175 & 113 & 157 & 164 \\
  \hline
\end{tabular}
\end{table}

\begin{figure}[h]
    \centering
    \includegraphics[width=0.475\textwidth]{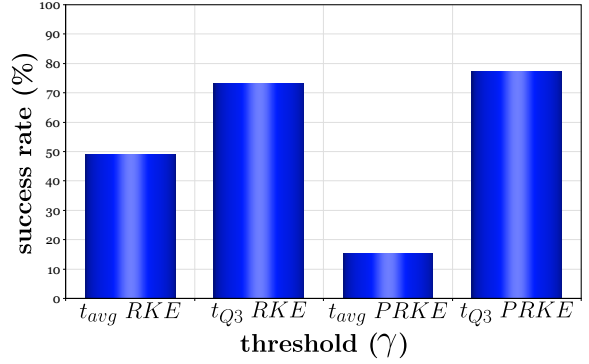}
    \caption{Success rate when trying to execute a command in both RKE and PRKE systems considering different thresholds.}
    \label{fig:usability}
\end{figure}

At this point, we could think on the possibility of choosing the maximum value as the threshold. However, it could be dangerous if a relay attack is performed: for the RKE system, if the message takes the minimum time $t_{min} = 55 \ ms$ to go to the attacker $A_1$, and the second attacker $A_2$ gets to send the relayed message in the same amount of time, it would take $110 \ ms$. Assuming that the attackers will not be able to exchange the relayed message through a LTE network or similar in less than $t_{max} - 110 = 26 \ ms$ is a weak premise. To solve this, we could take the average amount of time, but then we are compromising the usability of the system, so most of the time the user will have to press the button more than once, as shown in Figure \ref{fig:usability}. We can overcome this problem by calculating the third quartile of the dataset, which is higher than the average in both RKE and PRKE systems. We can see in Figure \ref{fig:usability} that now the effectivity is higher as well. As every time we press the button in the fob we are sending around 6 messages, the probability of failing when trying to execute a command is almost negligible, so the success rate for each message is almost 75\%.

\subsection{Robustness against relay attacks}

\begin{figure}[h]
    \centering
    \includegraphics[width=0.475\textwidth]{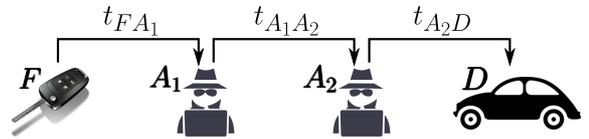}
    \caption{RKE relay attack scenario.}
    \label{fig:relayTime}
\end{figure}

Let us have an RKE relay attack scenario as depicted in Figure \ref{fig:relayTime}. If the minimum time it can ever take for the user's hardware to send a message from \textit{F} to \textit{D} is $t_{min}$, we can be sure that $t_{FA_1} = t_{min}$ is the minimum value that can be achieved. As such, our scheme is secure as far as the attackers are not able to achieve the following statement:
\begin{equation}\label{eq1}
\begin{split}
t_{FA_1} + t_{A_1A_2} + t_{A_2D} \leq \gamma \\
(t_{A_1A_2} + t_{A_2D}) \leq \gamma - t_{FA_1} \\
(t_{A_1A_2} + t_{A_2D}) \leq \gamma - t_{min}
\end{split}
\end{equation}

\begin{figure}[h]
    \centering
    \includegraphics[width=0.475\textwidth]{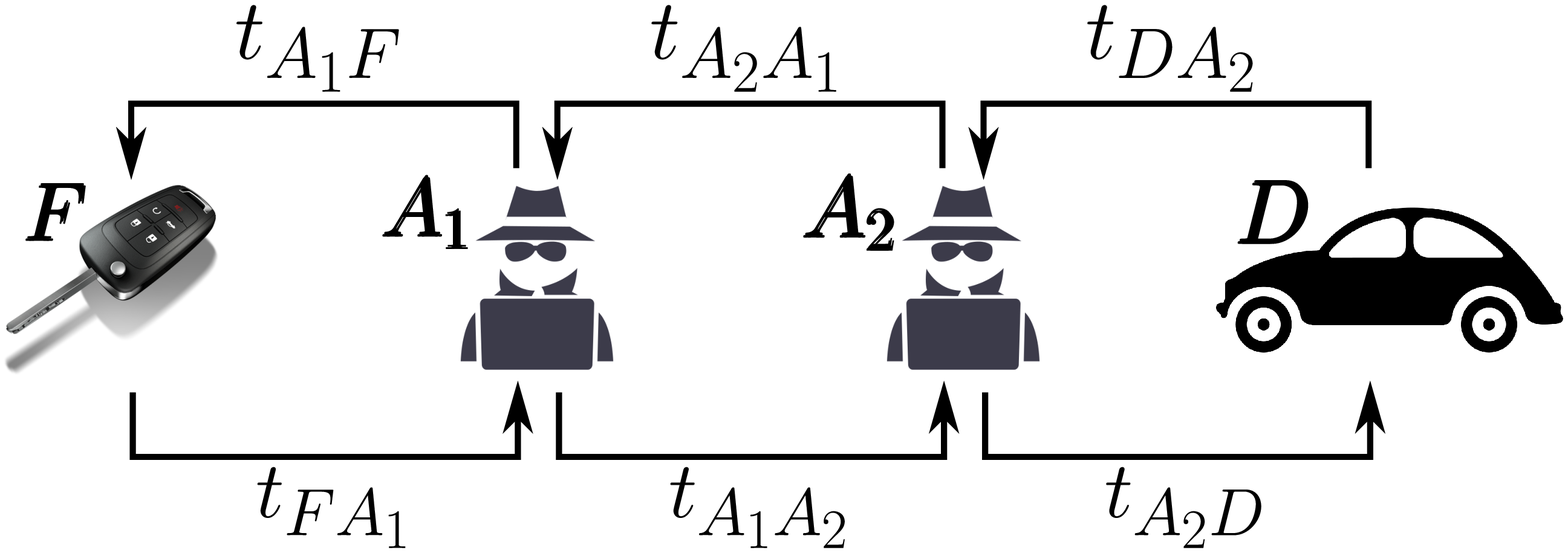}
    \caption{PRKE relay attack scenario.}
    \label{fig:prelayTime}
\end{figure}

On the other hand, we have a PRKE relay attack scenario as depicted in Figure \ref{fig:prelayTime}. If the minimum time it can ever take for the user's hardware to send a message from \textit{D} to \textit{F} and send the answer back to \textit{D} is $t_{min}$, we can be sure that $(t_{DA_2} + t_{FA_1}) = t_{min}$ is the minimum value that can be achieved. As such, our scheme is secure as far as the attackers are not able to achieve the following statement: 
\begin{equation}\label{eq2}
\begin{split}
t_{DA_2} + t_{A_2A_1} + t_{A_1F} + t_{FA_1} + t_{A_1A_2} + t_{A_2D} \leq \gamma \\
(t_{A_2A_1} + t_{A_1F} + t_{A_1A_2} + t_{A_2D}) \leq \gamma - (t_{DA_2} + t_{FA_1}) \\
(t_{A_2A_1} + t_{A_1F} + t_{A_1A_2} + t_{A_2D}) \leq \gamma - t_{min}
\end{split}
\end{equation}

If we take as an example the results we got, the attackers trying to hack \textit{LASER} should achieve the next statements to succeed, where $\gamma = t_{Q3} = 79 \ ms$ and $t_{min} = 55 \ ms$ for RKE:
\begin{equation}\label{eq3}
(t_{A_1A_2} + t_{A_2D}) \leq 24 \ ms
\end{equation}

And $\gamma = t_{Q3} = 164 \ ms$ and $t_{min} = 113 \ ms$ for PRKE:
\begin{equation}\label{eq4}
(t_{A_2A_1} + t_{A_1F} + t_{A_1A_2} + t_{A_2D}) \leq 51 \ ms
\end{equation}

As detailed in Section \ref{sec:background}, the bridge between $A_1$ and $A_2$ can be done through an LTE network or similar. Knowing that the average uplink latency in LTE networks is $10.5 \ ms$ \cite{latency}, we could assume two attackers getting lower values for $t_{A_1A_2}$ and $t_{A_2A_1}$. Even so, assuming that a relay attack can be successful against \textit{LASER} is a strong premise.

\section{\uppercase{Conclusion}}
\label{sec:conclusion}
\noindent
In this paper we have introduced \textit{LASER}, a lightweight and secure scheme for both RKE and PRKE systems. \textit{LASER} solves the security issues present into these systems, completely avoiding jamming-and-replay and relay attacks without using complex cryptographic schemes. Furthermore, it mitigates DoS attacks thanks to a simple frequency-hopping protocol. \textit{LASER} is easy-to-implement and we demonstrated it by implementing a prototype using non-expensive hardware. Last but not least, we proved the effectiveness and robustness of our solution through different experiments we performed.

\section*{ACKNOWLEDGEMENTS}
This work was supported by Project RTI2018-102112-B-I00 and by the European Research Council under the H2020 Framework Programme/ERC grant agreement 694974.

\bibliographystyle{apalike}
{\small
\bibliography{bibtex}}

\end{document}